\documentclass[12pt,a4]{article}
\usepackage{graphicx}
\usepackage{longtable}
\usepackage{amsmath}
\usepackage[all]{xy}

\newcommand{\be}{\begin{equation}}
\newcommand{\ee}{\end{equation}}
\newcommand{\ba}{\begin{eqnarray}}
\newcommand{\ea}{\end{eqnarray}}
\newcommand{\baa}{\begin{eqnarray*}}
\newcommand{\eaa}{\end{eqnarray*}}

\begin{document}

\title{Emission and absorption noise in the fractional quantum Hall effect}
\author{Cristina Bena$^{1,2}$ and In\`es Safi$^2$\\
{\small \it 1 Service de Physique Th\'eorique, CEA/Saclay},
\vspace{-.1in}\\{\small \it  Orme des Merisiers, 91190 Gif-sur-Yvette CEDEX}
\\{\small \it 2 Laboratoire de Physique des Solides},
\vspace{-.1in}\\{\small \it Universit\'e Paris-Sud, 91405 Orsay CEDEX}}
\maketitle

\begin{abstract}

We compute the high-frequency emission and absorption noise in a fractional quantum Hall
effect (FQHE) sample at arbitrary temperature.
We model the edges of the FQHE as chiral Luttinger liquids
(LL) and we use the non-equilibrium perturbative Keldysh
formalism. We find that the non-symmetrized high frequency noise
contains important signatures of the electron-electron
interactions that can be used to test the Luttinger liquid
physics, not only in FQHE edge states,
but possibly also in other one-dimensional systems such as carbon
nanotubes. In particular we find that the emission and absorption
components of the excess noise (defined as the difference between the noise at
finite voltage and at zero voltage) are different in an interacting
system, as opposed to the non-interacting case when they are
identical. We study the resonance features which appear in the
noise at the Josephson frequency (proportional to the applied voltage),
and we also analyze the effect of the
distance between the measurement point and the backscattering site.
Most of our analysis is performed in the weak backscattering limit, but we also
compute and discuss briefly the high-frequency noise
in the tunneling regime.

\end{abstract}

\section{Introduction}

In spite of intense theoretical and experimental exploration over
the past years, many features of one-dimensional strongly interacting systems have
not yet been clarified. For example, charge fractionalization, which has been observed in fractional quantum Hall effect(FQHE) edge states\cite{glattli}, has not been observed directly in carbon nanotubes. Furthermore, though it is believed
\cite{shotproposal,safiprl01,eunah,chetan} that the statistics of LL quasiparticles can be
measured in various setups, actual experiments addressing this
issue are much fewer \cite{statexp}.
Shot noise is a powerful tool to extract information about the  charge and statistics of the elementary excitations of a system. In particular, it is believed that high frequency noise contains important information about the statistics of quasiparticles \cite{eunah,chetan}, as well as about the charge fractionalization in systems such as carbon nanotubes, where charge fractionalization is masked at zero frequency by the metallic leads \cite{dolcini,ines}.
While some high frequency noise measurements have been performed for diffusive conductors \cite{sch}, the relevant range of frequencies was until recently too high and out of the experimental reach for one-dimensional systems; with new experiments being developed \cite{deblock,glattli-new}, this range of frequencies will likely become experimentally accessible in the near future. Moreover, these groups will have access to the non-symmetrized noise, i.e. both the emission and the absorption portions of the noise. The emission noise (the non-symmetrized noise at positive frequencies) quantifies the amount of photons of specific frequencies which are emitted by the fluctuating system, while the absorption noise (the non-symmetrized noise at negative frequencies) measures the amount of photons emitted by an active detector that can be absorbed by the fluctuating system. In general the emission noise is non-zero only when energy is put into the system (e.g. by applying an external voltage, or working at finite temperature), and only for frequencies smaller than the applied voltage/temperature, consistent with energy conservation; however there is no such constraint on the absorption noise.

High-frequency symmetrized noise at zero temperature has been addressed theoretically \cite{chamon1,saleur,ffnoise}. Similarly, some aspects of the non-symmetrized noise have been addressed theoretically for non-interacting systems \cite{deblock} and for situations in which the effects of interactions can be taken into account perturbatively \cite{hekking}. However, non-symmetrized noise in one-dimensional interacting systems has not been previously analyzed theoretically.

Here we compute the high-frequency finite-temperature non-symmetrized noise in a FQHE sample at filling factors of a simple fraction $\nu=1/2n+1$ form, and we compare the results to the symmetrized noise. Most of our analysis is done for the weak backscattering limit, but we also compute and discuss briefly the opposite strong-backscattering limit.

\begin{figure}[htbp]
\begin{center}
\includegraphics[width=3in]{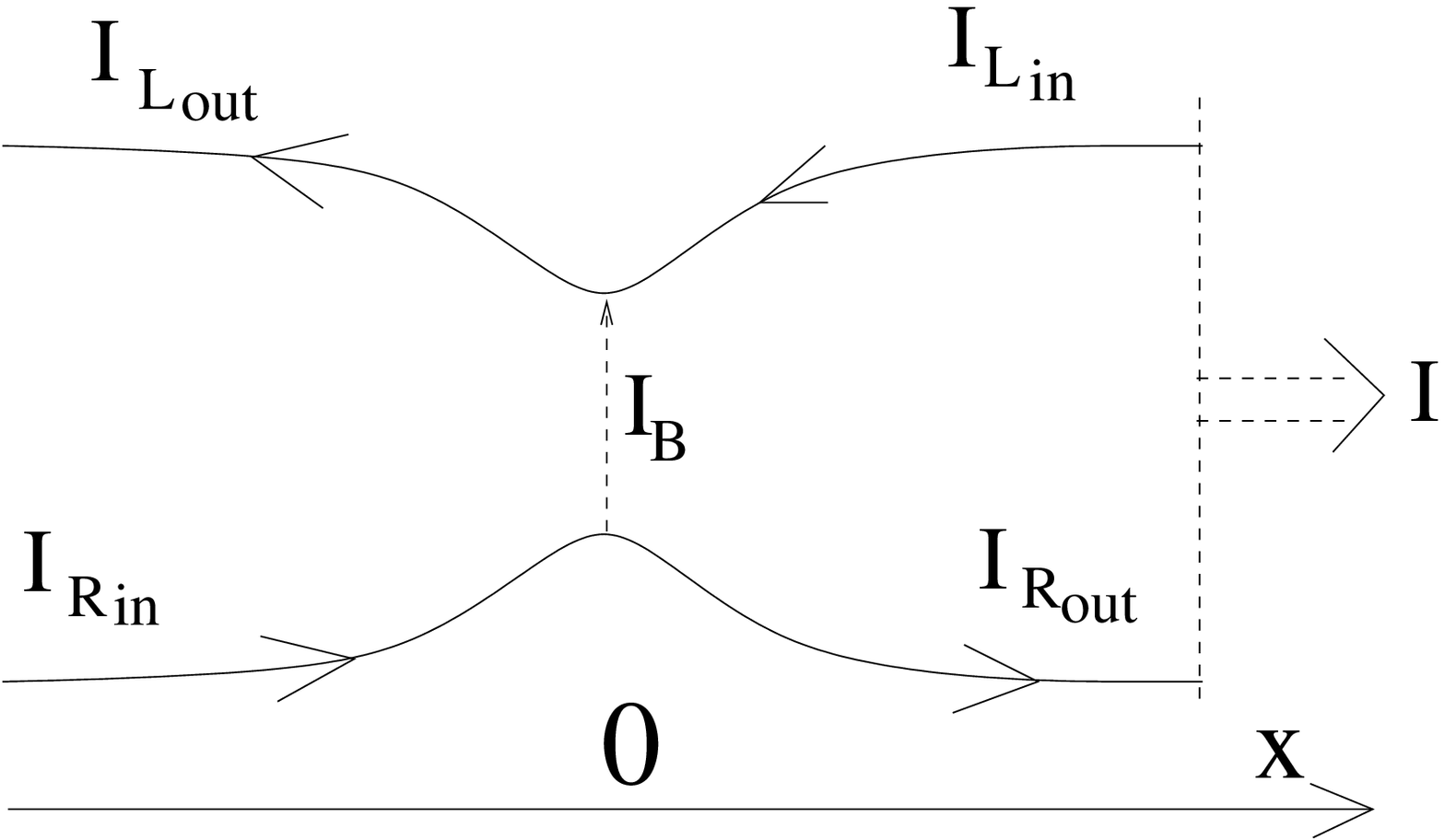}
\vspace{0.15in} \caption{\small Schematics of the sample}
\label{sample}
\end{center}
\end{figure}

The symmetrized noise is proportional to the Fourier transform of the expectation value of the anticommutator of two current fluctuation operators $\langle \{\Delta j(y,0),\Delta j(x,t)\} \rangle$; the non-symmetrized noise is proportional to the Fourier transform of $\langle \Delta j(y,0) \Delta j(x,t) \rangle$.
We should stress that we can obtain access to both the quantum
regime when the frequency is much larger than the temperature, and the classical regime, in which the frequency is much smaller than the temperature. We focus on \\
$\bullet$ the auto-correlations (noise) in the total current flowing trough the system,
\\
$\bullet$ auto-correlations and cross-correlations of individual currents flowing trough the four terminals of a FQHE sample (see Fig.~\ref{sample}).

The auto-correlations of the total current, as well as of individual branches contain two types of noise:
\\
$\bullet$ noise in the absence of backscattering  (which is also independent of voltage).
\\
$\bullet$ backscattering-induced noise (which is the difference between the noise in
the presence and in the absence of backscattering).

The cross-correlations between the outgoing right-movers and the
outgoing left-movers are solely backscattering-induced. In our
analysis we will focus mainly on the backscattering-induced
component of the noise, as it probes the very nature of
quasiparticles.

When we study backscattering-induced noise we also need to
distinguish between
\\
$\bullet$ zero voltage noise,
\\
$\bullet$ the excess noise, which is the difference between the noise at a finite voltage and the noise at zero voltage. Since the noise in the absence of backscattering is independent of voltage, this is the same as the difference between the backscattering-induced noise at a finite voltage and the backscattering-induced noise at zero voltage.

We focus first on the auto-correlations of chiral outgoing
branches, and the cross-correlations between outgoing branches
with opposite chiralities. We find that these are entirely
even in frequency, and also do not depend on the distance
between the point where the current is measured and the
backscattering site. The most striking feature observed is a
singularity at the Josephson frequency (JF) $\omega_0=\nu e V/\hbar$
(proportional to the applied voltage $V$, and to
the filling factor $\nu$ equal to the fractional charge $g$). This feature appears as a
cusp in a non-interacting system ($\nu=1$) such that the noise
decreases linearly from $2 g e I_B$ at zero frequency to zero at
the JF. For interacting systems with $g<1/2$, for example for
$g=\nu=1/3$ the noise exhibits an inverse power-law divergence at
the JF, and decays to zero for frequencies larger than the JF. At
finite temperature the singularity is rounded-off, and the noise
exhibits a peak slightly below the JF. The position and the width
of the peak, as well as the manner in which the noise decays to
zero above the JF depend on temperature: at low
temperature the peak is very sharp and close to the JF, but when
the temperature is increased the peak moves to lower frequencies,
widens and disappears completely for temperatures comparable to
the applied voltage. The zero temperature limit of our results is consistent
with the behavior of the symmetrized cross-correlations obtained in Ref. \cite{chamon1}.

Next we analyze the dependence on frequency of the non-symmetrized
noise in the total current. We focus first on the
situation in which the distance between the point where the
current is measured and the backscattering site is much smaller than
$v_F /\nu \omega$, where $v_F$ is the Fermi velocity, for all
frequencies $\omega$ probed experimentally. We find that the
emission noise is roughly equal to the noise in the outgoing
chiral branches, however the absorption noise  exhibits a positive
peak slightly below the JF, and a negative dip slightly above the
JF. The average of the two, which is the symmetrized noise, is
similar in structure to the absorption noise.

Another important quantity that we study is the non-symmetrized excess noise (the difference between the non-symmetrized noise at finite and zero voltage). Consistent with previous findings \cite{deblock} we show that this is even for non-interacting electrons. However we find that it becomes non-symmetric in the presence of interactions. This is a signature of Luttinger liquid physics which will probably exist also in other one-dimensional systems such as carbon nanotubes even in the presence of metallic leads.

If the distance $x$ between the backscattering site and the measuring point
is significant, oscillations with a
period $2 \pi v_F/ \nu x$ in the noise dependence on frequency  also appear. Such oscillations do
not occur in the auto-correlations of individual branches or in
cross-correlations of outgoing branches, but are manifest in the
absorption noise and in the symmetrized noise of the total current. If an
average over the position of the measuring point is performed,
both the non-symmetrized and the symmetrized noise are reduced to
the form of the individual branch correlations described
previously.

We treat the edges of the FQHE as infinite chiral Luttinger liquids, and we use a perturbative Schwinger-Keldysh formalism \cite{Keldysh} to compute the backscattered current and the noise up to second order in the backscattering amplitude.
 In section 2 we present the basics of the mathematical formalism used to calculate the noise. Some of the details of the calculation are outlined in Appendices A and B. In Section 3 we present our results for the non-symmetrized noise and a comparison with the symmetrized noise. In section 4 we discuss our results in comparison with the Landauer-B\"uttiker approach, we discuss also the correspondence with the generalized Kubo formula and with the fluctuation-dissipation theorem, as well as the results for the high-frequency non-symmetrized noise in the tunneling limit. We conclude in section 5.

\section{Formalism}

We model the edges of the FQHE as an infinite Luttinger liquid described by the Hamiltonian
\begin{equation}
{\mathcal{H}} ={\mathcal{H}}_{0}  \, + \, {\mathcal{H}}_{B}, \label{L}
\end{equation}
where ${\mathcal{H}}_{0}$ describes the interacting one-dimensional system, and ${\mathcal{H}}_{B}$ accounts for the backscattering.
Explicitly, we have
\begin{eqnarray}
{\mathcal{H}}_0 &=&\frac{\hbar v_F}{2}  \int_{-\infty}^{\infty}
 dx \left[ \Pi^2 + \frac{1}{g^2}
(\partial _x\Phi )^2\right] \\
{\mathcal{H}}_B &=& \Gamma e^{i \sqrt{4 \pi} \Phi(x_0,t)+2 i k_F
x_0}+ h.c.
\label{l0}
\end{eqnarray}
Here, $\Phi(x,t)=\Phi_R(x,t)+\Phi_L(x,t)$ is the standard Bose field operator in
bosonization and $\Pi(x,t)=\partial_t \Phi(x,t)/v_F$ is its conjugate momentum density.
The parameter $g$ is the value of the fractional charge of the free excitations of the model; for the FQHE it has been shown \cite{fqhe} that $g$ is equal to the filling factor $\nu$. Also $k_F$ is the Fermi wavelength, and $x_0$ is the position of the backscattering site; for simplicity we will set $x_0$ to zero in our calculations.

In bosonization, the current operators are related to the bosonic
chiral fields $\Phi_{R/L}$ through
\begin{equation}
j_{R/L} (x,t) = \frac{e}{\sqrt{\pi}} \partial_t
\Phi_{R/L}(x,t) \;  \label{current} \\
\end{equation}
and
\be
j(x,t)=j_R(x,t)+j_L(x,t).
\ee
The presence of an applied voltage can be taken into account by the shift in the tunneling operator such that $\Gamma\rightarrow\Gamma e^{i \omega_0 t}$, where $\omega_0=g e V/\hbar$ is the Josephson frequency associated with the applied voltage.

The finite frequency non-symmetrized noise $S^1_{a b}(x,y,\omega)$ is defined as
\begin{eqnarray}
S^1_{a b}(x,y,\omega) \equiv 2 \int_{-\infty}^{\infty} dt~e^{i\omega t}
\left\langle \Delta j_a (y,0)  \Delta j_b(x,t)
\right\rangle \; , \label{noise}
\end{eqnarray}
where $a,b$ stand for the right/left moving indices, and $\Delta
j_a(x,t) = j_a(x,t) - \langle j_a(x,t) \rangle$ is the current
fluctuation operator at position $x$ and time $t$.
For positive frequencies, Eq. (\ref{noise}) corresponds to emission noise, and for negative frequencies it corresponds to absorption noise. The noise in the total current is given by
\be
S^1(x,y,\omega)= S^1_{RR}(x,y,\omega)+S^1_{LL}(x,y,\omega)+S^1_{LR}(x,y,\omega)+S^1_{RL}(x,y,\omega).
\ee
The symmetrized noise can be obtained by symmetrizing the above with respect to frequency, i.e.
\be
S^{0}_{a b}(x,y,\omega)=[S^1_{a b}(x,y,\omega) +S^1_{ba}(y,x,-\omega)]/2,
\ee
and similarly for the noise in the total current.
The calculation of the symmetrized and of the non-symmetrized noise is performed using the non-equilibrium Schwinger-Keldysh formalism\cite{Keldysh}; the details are presented in Appendices A and B.

\section{Results}
We find the noise in the absence of backscattering to be
\ba
s^{1}_{a b}(x,y,\omega)&=&
\delta_{a b}\frac{2 e^2 \omega^2}{\pi}\tilde{\cal C}^{+-}_{a}(x,y,\omega)\label{s01}
\\
s^{0}_{a b}(x,y,\omega)&=&
\delta_{a b}\frac{2 e^2 \omega^2}{\pi}\tilde{\cal C}^{\cal K}_{a}(x,y,\omega)
\ea
or equivalently, using the Keldysh transformations presented in the Appendices,
\ba
s^{\alpha}_{a b}(x,y,\omega)
&=&\delta_{a b}\frac{e^2 \omega^2}{\pi} \{\tilde{\cal C}^{\cal K}_{a}(x,y,\omega)+\alpha [\tilde{\cal C}^{\cal A}_a(x,y,\omega)-
\tilde{\cal C}_a^{\cal R}(x,y,\omega)]\}
\label{seq}
\ea
Since our formulas will apply both for symmetrized and non-symmetrized noise, we will use the index $\alpha$ to distinguish between them, such that $\alpha=0$ for
the symmetrized noise and $\alpha=1$ for the non-symmetrized noise. Also $\tilde{\cal C}^{+-}$ is the Fourier transform of the two point correlator which is given by ${\cal C}^{+-}_{a b}(x,0;y,t)= \langle \Phi_a(y,t) \Phi_b(x,0) \rangle$. We can see that Eq.(\ref{s01}) is consistent with
the definitions from Eq.(\ref{current}).
The $\tilde{\cal C}^{{\cal R},{\cal A},{\cal K}}$ are the corresponding Fourier transforms of the retarded, advanced and symmetric Green's functions of the system in the absence of backscattering. They are presented in detail in Appendix B, and we can write
\ba
&&\tilde{\mathcal{C}}^{\cal A}_{L/R}(x,y,\omega)=-\frac{g}{2 \omega}e^{\mp i g \omega (x-y)/v_F} \theta[\pm (x-y)] \nonumber \\&&
\tilde{\mathcal{C}}^{\cal R}_{L/R}(x,y,\omega)=\frac{g}{2 \omega}e^{\mp i g \omega (x-y)/v_F} \theta[\mp (x-y)] \nonumber \\&&
\tilde{\mathcal{C}}^{\cal K}_{L/R}(x,y,\omega)=\frac{g}{2 \omega}e^{\mp i g \omega (x-y)/v_F} \coth\Big(\frac{\hbar \omega}{2 k_B T}\Big).
\label{gf}
\ea
For simplicity we present two situations, $x=y>0$ (the currents
are measured at the same position), and $x=-y>0$ (the currents are
measured at equal and opposite distances from the backscattering site). More general
situations can be easily obtained from the formalism presented in
the Appendices.
For $x=y>0$ we can write for the noise in the absence of backscattering:
\ba
&&s^{\alpha}_{R R}(x,x,\omega)=s^{\alpha}_{L
L}(x,x,\omega)=\frac{g \omega e^2}{2 \pi}
\Big[\coth\Big(\frac{\hbar \omega}{2 k_B T}\Big)-\alpha\Big] \ea
for the chiral branches, with the cross-correlations
between the right and the left-movers vanishing in the absence of backscattering: $s^{\alpha}_{R
L}(x,-x,\omega)=0$. For the total current we find \ba
&&s^{\alpha}(x,x,\omega)=\frac{g \omega e^2}{\pi}
\Big[\coth\Big(\frac{\hbar \omega}{2 k_B T}\Big)-\alpha\Big].
\label{seqt} \ea

As also described in the Introduction, the total
noise is the sum of  the noise in the absence of backscattering and of
the backscattering-induced noise: $S^{\alpha}(x,y,\omega)=s^{\alpha}(x,y,\omega)+\delta
S^{\alpha}(x,y,\omega)$.
Following the treatment described in Appendix A, using the same notations and following similar lines with Ref. \cite{dolcini},
we find that
the backscattering-induced components of the correlations between
chiral currents can be written as:
\be
\delta S^{\alpha}_{a b}
(x,y,\omega)=S^A_{ a b}(x,y,\omega)+S^C_{a b}(x,y,\omega)+\alpha S^N_{a b}(x,y,\omega),
\label{qw}
\ee
where again $\alpha=0$ for the symmetrized noise and $\alpha=1$ for the non-symmetrized noise.
Here we have
\ba
S^A_{a b}(x,y,\omega)&=&-\frac{\omega^2}{\pi} \tilde{\cal C}^{\cal R}_{a}(x,0,\omega)
\tilde{\cal C}^{\cal R}_{b}(y,0,-\omega)f_A(\omega),\label{n1} \\
S^C_{a b}(x,y,\omega)&=&-\frac{\omega^2}{\pi}[ \tilde{\cal C}^{\cal K}_{a}
(x,0,\omega)\tilde{\cal C}^{\cal R}_{b}(y,0,-\omega)f_C(-\omega)
\nonumber \\&&
+\tilde{\cal C}^{\cal R}_{a}(x,0,\omega)\tilde{\cal C}^{\cal K}_{b}
(y,0,-\omega)f_C(\omega)],\label{n2}\\
S^N_{a b}(x,y,\omega)&=&-\frac{\omega^2}{\pi}[\tilde{\cal C}^{\cal A}_{a}(x,0,\omega)
\tilde{\cal C}^{\cal R}_b(y,0,-\omega)f_C(-\omega)
\label{n3} \\&& -
\tilde{\cal C}^{\cal R}_a(x,0,\omega)\tilde{\cal C}^{\cal A}_b(y,0,-\omega)f_C(\omega)].
\nonumber
\ea

The functions $f_A(\omega)$ and $f_C(\omega)$ are given by
\begin{equation}
f_A(\omega) = \int_{-\infty}^\infty dt \, e^{i \omega t}
\left\langle \left\{ \Delta j_B(t), \Delta j_B(0) \right\}
\right\rangle , \label{s_A}
\end{equation}
where $\Delta j_B(t) = j_B(t) - \langle j_B(t)
\rangle$. The expectation values are performed with respect to the full action.
Also, $j_B(t)$ is the backscattering
current operator at the backscattering site
\begin{equation}
j_{B}(t) = - \frac{e}{\hbar} \frac{\delta
\mathcal{H}_B(\Phi)}{\delta \Phi(0,t)} \; \label{ib_def} .
\end{equation}
Similarly
\begin{equation}
f_C(\omega)= \int_0^\infty dt \left( e^{i \omega t}-1 \right)
\left\langle \left[ j_B(t),j_B(0) \right]
\right\rangle \; . \label{s_C}
\end{equation}

In this notation, the emission noise is the non-symmetrized noise taken at positive
frequencies, while the absorption noise is the non-symmetrized
noise at negative frequencies.

We note that by summing up the chiral components of the noise to obtain the noise in the total current,
we recover the same structure for the symmetrized noise as the one presented in Ref.\cite{dolcini}.

Up to this point the calculation is non-perturbative and we can analyze a few {\it non-perturbative}
features of our results.
For example we note that if one studies the case $x=y>0$,
the fluctuations in the chiral currents will be independent of position. Similarly
for the situation $x=-y>0$, the
cross-correlations between the outgoing right-movers and the
outgoing left-movers are spatially independent. This is because the noise terms described above depend on position only through the free propagation of a chiral
moving mode from the first measuring point to the backscattering
site $\tilde{\mathcal{C}}_{L/R}(x,0,\omega)$, and through the free
propagation of another chiral mode from the backscattering site
to the second measuring point $\tilde{\mathcal{C}}_{L/R}(y,0,-\omega)$.
Thus the total phase accumulated during the propagation cancels
for $\delta S_{RR/LL}(x,x,\omega)$ and for $\delta S_{RL}(x,-x,\omega)$ which are position
independent. However other quantities, such as
the cross-correlations between the incoming and outgoing right-movers $\delta S_{RR}(x,-x,\omega)$,
are affected by interference effects, and are hence position dependent.

We also note that, as it can be seen from the above
non-perturbative formulas and from Eqs.(\ref{gf}), the chiral $\delta S^1_{RR}(x,x,\omega)$, and $\delta
S^1_{RL}(x,-x,\omega)$ current correlations are equal to each other, and also even in frequency
(independent of $\alpha$ - the emission
noise is equal to the absorption noise). However $\delta
S^1_{RR}(x,-x,\omega)$ and the backscattering noise in the total
current $\delta S^1(x,x,\omega)$ are non-symmetric (dependent on $\alpha$ - the emission
noise is different from the absorption noise).

\subsection{Perturbative results}
Using Eq.(\ref{qw}) we can now evaluate our results perturbatively up to second order in $\Gamma$ to find
\ba &&\delta S^{\alpha}_{R
R}(x,x,\omega)= \delta S^{\alpha}_{R L}(x,-x,\omega) \nonumber
\\&& =g e \sum_{m=\pm 1}\Big\{\coth\Big[\frac{\hbar (\omega+m
\omega_0)}{2 k_B T}\Big]-\coth\Big(\frac{\hbar \omega}{2 k_B
T}\Big)\Big\} I_B(\omega+m  \omega_0). \label{crosseq1} \ea
Similarly
\be \delta S^{\alpha}_{R R/L L}(x,-x,\omega)= \mp
\frac{g^2}{4 \pi}\Big[\coth\Big(\frac{\hbar \omega}{2 k_B
T}\Big)-\alpha\Big]f_C(\pm \omega) e^{\pm 2 i g  \omega x/v_F}
\label{crosseq11} \ee
The other cross-correlators are evaluated in Appendix A in Eq.(\ref{qet}). Using the above formulas and
Eq.(\ref{qet}), we find the noise in the total current to be:
\ba
&&\delta S^{\alpha}(x,x,\omega)=g e \sum_{m=\pm
1}\Big\{\coth\Big[\frac{\hbar (\omega+m \omega_0)}{2 k_B
T}\Big]-\coth\Big(\frac{\hbar \omega}{2 k_B
T}\Big)\Big\}I_B(\omega+m  \omega_0)
 \\&&
-\Big[\coth\Big(\frac{\hbar \omega}{2 k_B T}\Big)-\alpha\Big]\Big\{
\sum_{m=\pm 1}I_B(\omega+m  \omega_0)\cos\Big( \frac{2 g  \omega x}{v_F}\Big)
+ i \frac{g^2}{4 \pi} \sin\Big(\frac{2 g  \omega x}{v_F}\Big)[f_C(\omega)+f_C(-\omega)]\Big\} \nonumber
\label{crosseq}
\ea

where we can evaluate $f_C$ (presented in Eq.(\ref{s_C})) perturbatively:
\ba
f_C(\omega)=4 \pi i \Big(\frac{e}{\hbar}\Big)^2 |\Gamma|^2 \int_0^\infty dt (e^{i \omega t}-1)
\cos(\omega_0 t) \Big[\frac{\pi t k_B T}{\hbar \sinh(\pi t k_B T/\hbar)}\Big]^{2 g} {\rm{Im}}\big[(1+i t \epsilon_h/\hbar)^{-2 g}\big].
\ea
Also
\be
I_B(\Omega)= - \frac{g e}{\hbar^2} |\Gamma|^2  {\cal{F}}_g(\Omega)
\ee
is the value of the backscattered current for an applied voltage equal to $\Omega$, with
\ba
&&{\cal F}_g(\omega)=\int_0^{\infty} \sin(\omega t)\Big[\frac{\pi t k_B T}{\hbar \sinh(\pi t k_B T/\hbar)}\Big]^{2 g} {\rm{Im}}\big[(1+i t \epsilon_h/\hbar)^{-2 g}\big] dt.
\label{calf}
\ea
When taking into account $\hbar\omega\ll\epsilon_h$ this becomes
\ba
&&{\cal F}_g(\omega) \approx -\sin(\pi g)\Big(\frac{\pi k_B T}{\epsilon_h}\Big)^{2 g} 2^{2 g} \int_0^{\infty} \sin(\omega t) \sinh^{-2 g}(\pi t k_B T/\hbar) dt
\nonumber \\
&&=i \sin(\pi g) \Big(\frac{\pi k_B T}{\epsilon_h}\Big)^{2 g} 2^{2 g-2} \Gamma(1-2 g)\Big[\frac{\Gamma(g-i \tilde{\omega})}{\Gamma(1-g-i \tilde{\omega})}-\frac{\Gamma(g+i \tilde{\omega})}{\Gamma(1-g+i \tilde{\omega})}\Big]
\ea
where $\tilde{\omega}=\hbar \omega/2 \pi k_B T$.

For $g=1/2$
\be
{\cal F}_{1/2}(\omega)=-\frac{\pi \hbar}{2 \epsilon_h} \tanh\Big(\frac{\hbar \omega}{2 k_B T}\Big)
\ee
while for $g=1$(non-interacting system),
\be
{\cal F}_1(\omega)=-\frac{\pi \hbar^2}{2 \epsilon_h^2} \omega
\ee

We can note directly from Eq. (\ref{crosseq}) that for frequencies much larger than the temperature, i.e in the quantum regime, the emission noise ($\omega >0$) is independent of position, and it decays to zero for frequencies slightly larger than the Josephson frequency.

We evaluate some simple limits ($g=1$, and $g=1/2$) to obtain:
\ba
\frac{\delta S^{\alpha}}{e I_B}&=&\frac{1}{\omega_0}\Big\{\sum_{m=\pm1}(\omega+m \omega_0)\coth\Big(\frac{\hbar \omega+m \hbar \omega_0}{2 k_B T}\Big)
-2\omega \Big[2\coth\Big(\frac{\hbar \omega}{2 k_B T}\Big)-\alpha\Big]\Big\}
\label{ni}
\ea
for $g=1$, where $I_B$ is the value of the backscattering current. The result for the symmetrized noise ($\alpha=0$) is in agreement with the Landauer-B\"uttiker formalism \cite{buttiker} which
finds that the noise dependence on frequency for an arbitrary amount of backscattering is:
\be
S^0(\omega)=\frac{e^2}{2 \pi \hbar} \Big[{\cal T}(1-{\cal T})\sum_{m=\pm1}(\hbar \omega+m e V)\coth\Big(\frac{\hbar \omega+m e V}{2 k_B T}\Big)+2 {\cal T}\hbar \omega \coth\Big(\frac{\hbar \omega}{2 k_B T}\Big) \Big],
\label{lb}
\ee
where $\cal T$ is the transmission of the barrier.
We see that by taking the limit ${\cal T}\rightarrow1$ (weak backscattering), and expanding this result in $1-{\cal T}\propto |\Gamma|^2$
we retrieve the same behavior as that presented in Eq.(\ref{ni}).

For $g=1/2$, we obtain
\be
\frac{\delta S^{\alpha}}{g e I_B}=\Big\{2-\Big[2 \coth\Big(\frac{\hbar \omega}{2 k_B T}\Big)-\alpha\Big] \sum_{m=\pm1}\tanh \Big(\frac{\hbar \omega+m\hbar \omega_0}{2 k_B T}\Big)\Big\}\coth\Big(\frac{\hbar \omega_0}{2 k_B T}\Big)
\ee
which for $\alpha=0$ (symmetrized noise) is consistent with the small $\Gamma$ limit of the exact calculation presented in Ref. \cite{chamon1} performed using a scattering approach.

We also note that the expansion of $I_B(x)$ around $x=0$ is of the form of $a x+ b x^3$.
Similarly, $I_B(x)\approx x^{2 g-1}$ for $x \gg 1$. This is consistent with the standard Luttinger liquid theory, where the current is linear with voltage for voltages much smaller than the temperature, and has a power-law dependence on voltage for voltages much larger than the temperature. Also, this implies that the zero-temperature inverse power-law divergences in the first term in $\delta S$ at the JF are rounded off quadratically at finite temperature. The second term in $\delta S$ is linear in the vicinity of the JF.

Before plotting our results we will also say a few words about the range of validity of the perturbative expansion. To ensure that the perturbative approach is valid we need to check that the backscattering-induced noise is much smaller than
the noise in the absence of backscattering. For frequencies far from the Josephson frequencies this translates into the criterion: $[max(e V, k_B T, \omega)/\epsilon_h]^{2-2 g} \gg |\Gamma|^2/\epsilon_h$, where $\epsilon_h$ is the high energy cutoff of the problem. For the regime when $\omega$ is comparable to $\omega_0$, the criterion is harder to write down, and we have to check that the height of the peaks near the Josephson frequency is sufficiently limited by the temperature.

\begin{figure}[htbp]
\begin{center}
\includegraphics[width=3in]{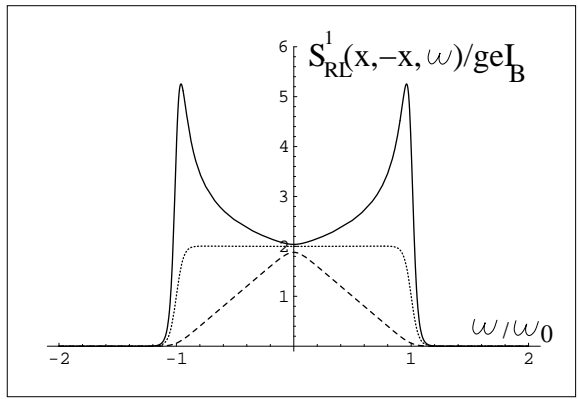}
\caption{\small The backscattering-induced {\it non-symmetrized} cross-correlations between the outgoing right-movers and the outgoing left-movers
renormalized by $ g e I_B$ plotted as a function of frequency (in
units of $\omega_0$) for $g=\nu=1/3$ (full line), $g=\nu=1$
(dashed line), and $g=1/2$ (dotted line). The $g=1/2$ situation
does not correspond to any FQHE state, but is drawn here only for
comparison. We set $k_B T/\hbar \omega_0=0.03$. Note that the cross-correlations are even (the emission component is equal to the absorption component). Note also that this plot describes also the excess cross-correlation noise (the zero voltage cross-correlations vanish), and that in the absence of backscattering the cross-correlations are zero.} \label{sym}
\end{center}
\end{figure}

In Fig.~\ref{sym} we plot the non-symmetrized
backscattering-induced correlations between the chiral currents
$\delta S^1_{R R}(x,x,\omega)=\delta S^1_{R L}(x,-x,\omega)$.
These correlations are $even$, and independent of position.
They also exhibit a singularity at the JF $\omega_0=g e V/\hbar$.
This is cusp-like for a non-interacting system ($\nu=1$). For
$g=1/2$ the noise has a step-like transition, while for $g=\nu=1/3$  the
noise exhibits a peak slightly below the JF, and falls to zero for
frequencies larger than the JF.

From Eq.(\ref{crosseq1}) we can see that in the absence of an applied voltage the correlators described above vanish (the backscattering-induced noise is equal to the excess noise). Also, while the fluctuations in the current of outgoing right-movers are non-zero in the absence of backscattering, there is no such component for the cross-correlations between the outgoing right-movers and the outgoing left-movers. Thus the cross-correlations are of great experimental relevance when we are interested in isolating the backscattering-induced noise; while for some of the noise quantities that can be measured in a FQHE four-terminal setup the contributions in the absence of backscattering (of order $1$) may mask the backscattering-induced noise (of order $|\Gamma|^2 \ll 1$), this is not the case for the cross-correlations.

\begin{figure}[htbp]
\begin{center}
\includegraphics[width=3in]{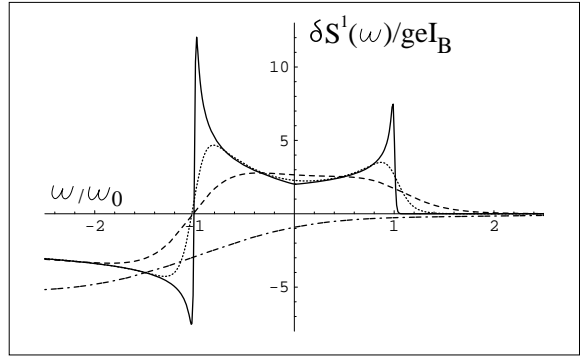}
\caption{\small The backscattering non-symmetrized noise in the total current $\delta S^1(x,x,\omega)$ renormalized
by $ g e I_B$ plotted as a function of frequency in units of
$\omega_0$ for $g=\nu=1/3$ and different temperatures $k_B T/\hbar \omega_0=0.01$(full line), $k_B T/\hbar \omega_0=0.1$ (dotted line), $k_B T/\hbar \omega_0=0.3$ (dashed line), and $k_B T/\hbar \omega_0=1$ (dashed-dotted line). We consider $g \omega x/v_F \ll 1$ and for simplicity we denote $\delta S^1(x,x,\omega)\equiv \delta S^1(\omega)$.}
\label{temp}
\end{center}
\end{figure}

We now present our results for the non-symmetrized noise in the total current.
For simplicity we start with the limit $g \omega x/v_F\ll1$ which is presented in Fig.~\ref{temp}. Consistent with Eq. (\ref{crosseq1}), we see that the emission noise is similar to the chiral current correlations presented above. At low temperature the emission noise vanishes for frequencies larger than the JF, consistent with the intuitive picture that no photon with an energy larger that the applied voltage can be emitted by the fluctuating system. While the emission noise behaves similar to the chiral current correlators, the absorption noise exhibits a positive peak slightly below the JF, and a negative dip slightly above the JF. The function connecting the negative and the positive resonances has a linear dependence of frequency when it passes through the JF.  As depicted in Fig.~\ref{temp}, with increasing the temperature the peaks broaden and move away from the JF, to disappear for temperatures of the same order of magnitude
as the applied voltage.

\begin{figure}[htbp]
\begin{center}
\includegraphics[width=3in]{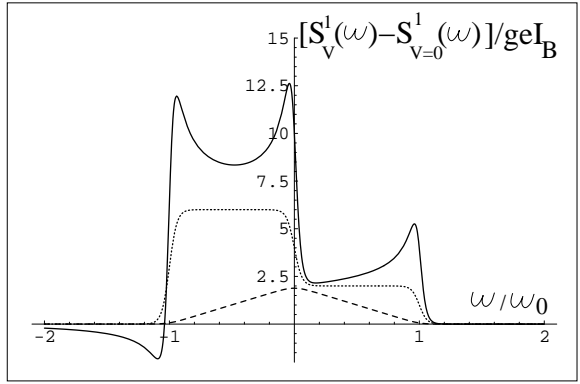}
\caption{\small The excess noise in the total current $S^1_V(x,x,\omega)-S^1_{V=0}(x,x,\omega)$ (in units of
$g e I_B$), as a function of frequency (in units of $\omega_0$) for $\nu=1$ (dashed line), $\nu=1/3$ (full line) and $\nu=1/2$ (dotted line). We set $k_B T/\hbar \omega_0=0.03$ and
$g \omega x/v_F\ll 1$. For simplicity we denote $S^1(x,x,\omega)\equiv S^1(\omega)$.}
\label{excessn}
\end{center}
\end{figure}

The corresponding excess noise is depicted in Fig.~\ref{excessn}. Indeed, as described in Ref. \cite{deblock}, the emission and absorption excess noises are equal for non-interacting electrons. However, they are {\it different} in the presence of interactions.
The asymmetry between the emission and absorption high-frequency excess noises is an important signature of Luttinger liquid physics which should be looked for  also in carbon nanotubes where the presence of metallic leads masks the physics of charge fractionalization in the zero frequency noise.
Also, we note that the absorption excess noise can even
become negative in some regions of frequency for $g<1/2$.
A similar behavior was also found
for the symmetrized noise of one-dimensional interacting systems connected to metallic leads \cite{dolcini}, as well as for the case of a ballistic single-channel
quantum wire, capacitively coupled to a gate \cite{negexcess}.

\begin{figure}[htbp]
\begin{center}
\includegraphics[width=3in]{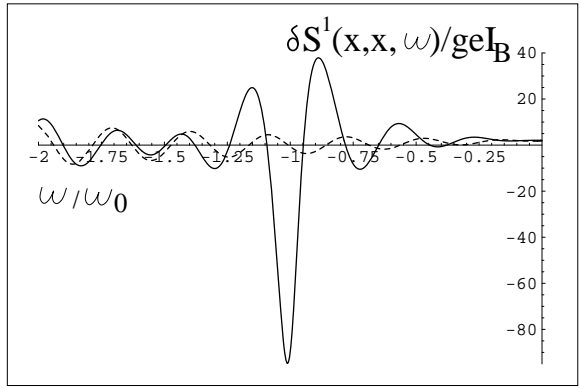}
\vspace{0.15in} \caption{\small The backscattering
absorption noise in the total current $\delta S^1(x,x,\omega<0)$ (renormalized by $ g e I_B$) as a function of frequency (in units of $\omega_0$) for $g=\nu=1/3$ (full line),
$g=\nu=1$ (dashed line). We set $k_B T/\hbar \omega_0=0.03$. For positive frequencies (emission noise), we retrieve the same behavior as the one described in Fig.~\ref{temp}.} \label{posn}
\end{center}
\end{figure}

The dependence of the absorption noise on frequency in the limit in which $g \omega x/v_F$ is of order $1$ is presented in Fig.~\ref{posn}. For the temperature considered here,
$k_B T/\hbar \omega_0=0.03$, the emission noise is basically unchanged and is still characterized by the behavior presented in Fig.~\ref{temp}; note that oscillations with a period of $\pi v_F/g x$ appear in the absorption part of the noise. By averaging over the position of the measuring point, the absorption noise in the total current actually becomes equal to the emission noise.
%

\begin{figure}[htbp]
\begin{center}
\includegraphics[width=3in]{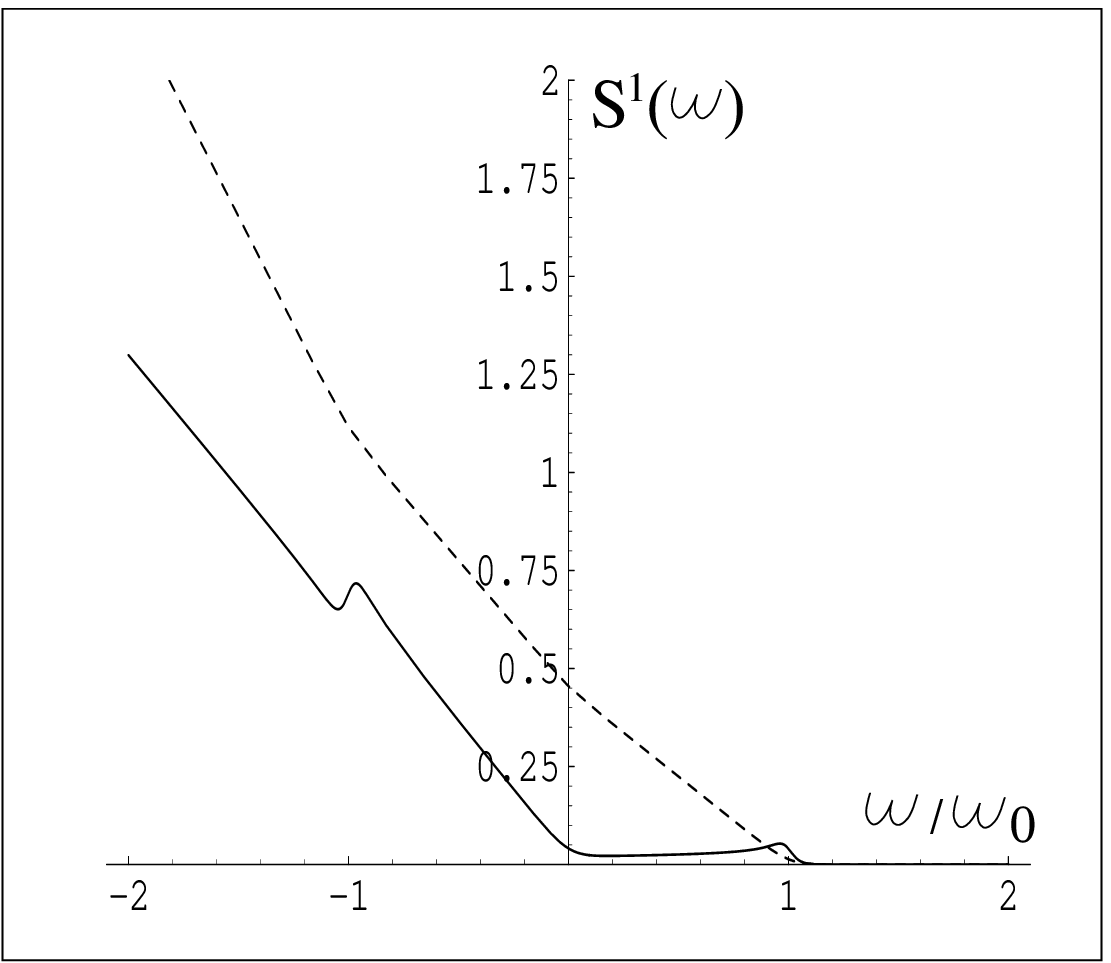}
\caption{\small The total non-symmetrized noise $S^1(x,x,\omega)=s^1(x,x,\omega)+\delta S^1(x,x,\omega)$ in the total current (including the noise in the absence of backscattering) for $g=1$ (dashed line) and $g=1/3$ (full line) (in arbitrary units), as a function of frequency (in units of $\omega_0$), for
$g \omega x/v_F\ll 1$. We set $k_B T/\hbar \omega_0=0.03$. For simplicity we denote
$S^1(x,x,\omega)\equiv S^1(\omega)$.}
\label{nstot}
\end{center}
\end{figure}

The total non-symmetrized noise (including the noise in the absence of backscattering) described by Eq.(\ref{seq}) is depicted in Fig.~\ref{nstot}.
We see that the emission spectrum is solely backscattering-induced,
consistent with the fact that at small temperature the amount of
photons that can be emitted by the system is very small if no
energy  is transferred to the system (e.g. through applying a
voltage).

\begin{figure}[htbp]
\begin{center}
\includegraphics[width=3in]{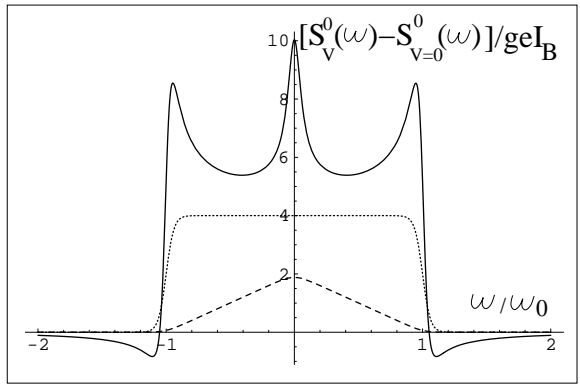}
\caption{\small The symmetrized excess noise in the total current $S^0_V(x,x,\omega)-S^0_{V=0}(x,x,\omega)$ in units of $g e I_B$, as a
function of frequency in units of $\omega_0$ for $g=\nu=1/3$ (full
line), $g=\nu=1$ (dashed line), and $g=1/2$ (dotted line). We set $k_B T/\hbar \omega_0=0.03$ and $g \omega x/v_F \ll 1$. For simplicity we denote
$S^0(x,x,\omega)\equiv S^0(\omega)$.}
\label{excess}
\end{center}
\end{figure}

Next we focus on the symmetrized noise. We note that in general it is not identical, but it has a similar behavior to the absorption noise. In Fig.~\ref{excess} we plot the
symmetrized excess noise in the total current as a
function of frequency for a few values of the parameter $g$, when
$g \omega x/v_F \ll 1$.  For a strongly interacting one-dimensional system such as the one described by $g=1/3$, the excess symmetrized noise has regions in which it becomes negative, similar to the absorption excess noise. Experimentally this may be used as a the signature of Luttinger liquid physics in the presence of strong interactions between electrons.

\begin{figure}[htbp]
\begin{center}
\includegraphics[width=3.5in]{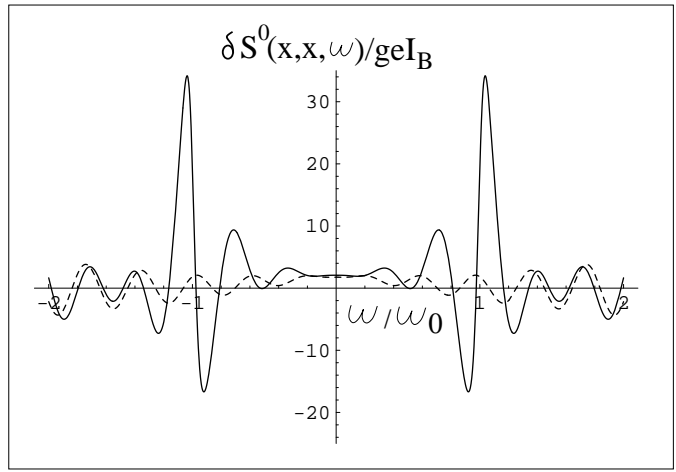}
\caption{\small The symmetrized backscattering-induced noise in the total current  $\delta S^0(x,x,\omega)$ (renormalized by $g e I_B$) as a function of frequency (in units of $\omega_0$) for
$\nu=g=1/3$ (full line), and $\nu=g=1$ (dashed line). The ratio
$\omega_0/\omega_L$ is set to $10$, where the energy scale
associated to the length $\omega_L= v_F/g x$.
We set $k_B T/\hbar \omega_0=0.03$.} \label{pos}
\end{center}
\end{figure}

When $g \omega x/v_F$ is of order $1$, the dependence of the symmetrized noise on frequency is depicted in Fig.~\ref{pos}. We note again the presence of oscillations with period $\pi v_F/g x$, similar to the case of the absorption noise. If an average with respect to the measuring point is performed, the symmetrized noise also reduces to the behavior of the cross-correlations presented in Fig.~\ref{sym}.

For the case of the total symmetrized noise the noise in the absence of backscattering (of order $1$) gives rise to a linear background with a slope proportional to $g$; backscattering (of order $|\Gamma|^2$) creates a small cusp-like feature at $\omega_0$ for $g=1$ and a small ``bump-like" feature for $g=1/3$. This is also similar to the absorption noise in Fig.~\ref{nstot}.

We would like to mention here that this behavior if similar to the behavior of a different physical system; as it turns out, one can map the impurity problem in a
Luttinger liquid to the problem of a coherent one-dimensional conductor embedded in an ohmic environment with an arbitrary resistance \cite{safisaleur}. For the latter, the high frequency symmetrized noise was computed
\cite{ffnoise} by combining a scattering matrix approach with a real time effective action formalism, and a similar behavior was found, with the difference that the Josephson singularity appears in that situation at a value proportional to the value of the voltage across the sample $U$ and not to the value of the voltage source $V$. We will examine the consequences of this mapping for the high-frequency non-symmetrized noise of a coherent conductor embedded in an ohmic environment in a separate publication.

\section{Discussion}

\subsection{Comparison with the Landauer-B\"uttiker approach for the symmetrized noise}
\begin{figure}[htbp]
\begin{center}
\includegraphics[width=3in]{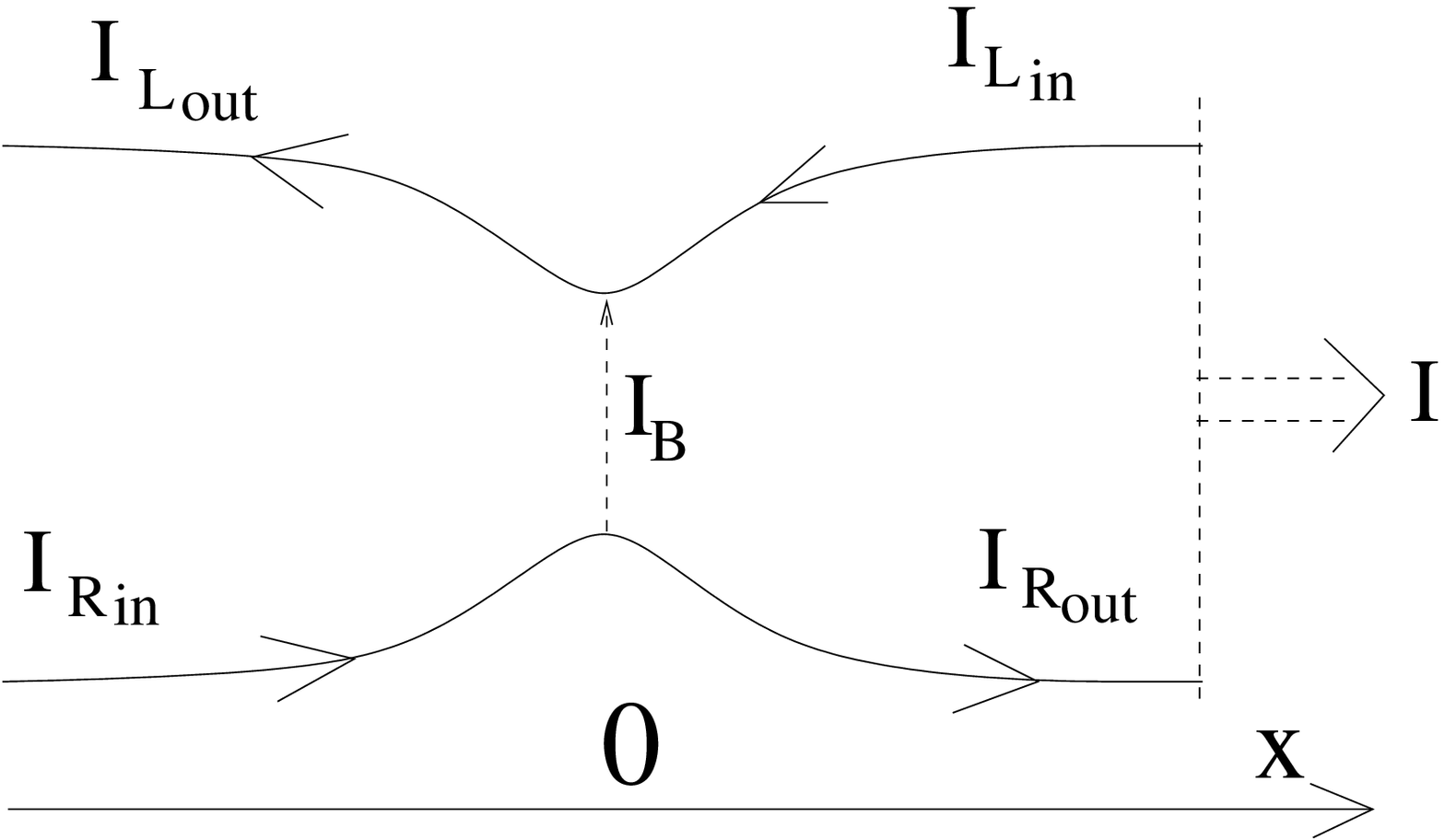}
\vspace{0.15in} \caption{\small Schematics of the sample}
\label{sample1}
\end{center}
\end{figure}

We can compare our results to those obtained using a  standard Landauer-B\"uttiker (LB)
approach \cite{buttiker} for non-interacting electrons.
In the limit when the distance from the backscattering site to the
measuring point is very small, $\omega x/v_F \ll 1$, we can write the symmetrized noise
in the total current \be S^0(t)=\langle
\{I_{R_{out}}(t)+I_{L_{in}}(t),I_{R_{out}}(0)+I_{L_{in}}(0)\}\rangle
\ee where $I_{R/L_{in/out}}$ are the currents of incoming/outgoing
right/left movers with the convention that all currents are
positive if flowing from left to right in Fig.~\ref{sample1}. The
tunneling current between the two edges is denoted $I_B$. Noting
that $I_{R_{in}}-I_{R_{out}}=I_{L_{in}}-I_{L_{out}}=I_B$, we can
rewrite the total noise (up to terms of the form
$\langle I_{R/L_{in}} I_{R/L_{in}}\rangle$) as \ba S^0(t)&=&\langle
\{I_B(t),I_B(0)\} \rangle+ \langle \{I_{R_{out}} (t),
I_{R_{in}}(0)\} \rangle + \langle \{I_{L_{out}} (t) ,I_{L_{in}}(0)
\}\rangle \nonumber \\&& +\langle \{I_{R_{in}} (t),
I_{R_{out}}(0)\} \rangle + \langle \{I_{L_{in}} (t)
,I_{L_{out}}(0) \}\rangle \ea

We can connect this relation to our results presented in
Eqs.(\ref{n1},\ref{n2}). We can identify the fluctuations in the
tunneling current between the edges $\langle \{I_B(t),I_B(0)\}
\rangle$ with the term $S_A$ in Eq.(\ref{n1}). As described in Eq.(\ref{lb}), for non-interacting
systems this term will be given by ${\cal T}(1-{\cal T})\sum_{m=\pm1}(\hbar \omega+m e V)\coth[(\hbar \omega+m e V)/2 k_B T]$. Perturbatively
we found this term to be $\sum_{m=\pm1}\coth[(\omega+m \omega_0)/2k_B T]$ $I_B(\omega+m
\omega_0)$. In the case
of small backscattering $1-{\cal T}\approx |\Gamma|^2$, and the two expressions are consistent in the non-interacting limit for which $I_B(\omega) \propto \omega$.

Similarly we see that we can identify the Fourier transform of $\langle \{I_{R_{out}} (t), I_{R_{in}}(0)\} \rangle$ \\
$+ \langle \{I_{L_{out}} (t), I_{L_{in}}(0) \}\rangle$ $+\langle \{I_{R_{in}} (t), I_{R_{out}}(0)\} \rangle $ $+ \langle \{I_{L_{in}} (t) ,I_{L_{out}}(0) \}\rangle$ with the term denoted $S_C$ in Eq.(\ref{n2}).
We note that this term is due to correlations between currents in the same reservoir.
In the LB formalism in the absence of interactions  (see Eq.(\ref{lb})) this term is proportional to ${\cal T}\hbar \omega$ $\coth(\hbar \omega/2 k_B T)$, while perturbatively
we found it to be proportional to $-\coth(\omega/2k_B T)$ $\sum_{m=\pm1}$ $I_B(\omega+m \omega_0)$.
We see that again the two expressions agree in the non-interacting limit for which $I_B(\omega) \propto \omega$, in the case of small backscattering.

In a four-terminal setup such as a FQHE bar, one has experimental access to the chiral current correlations as well as to the noise in the total current. We can see from the formulas presented in Eqs.(\ref{crosseq1},\ref{crosseq}) that in this situation one can also $measure$ the two terms described above separately, for example by combining the symmetrized cross-correlations and the symmetrized noise in the total current: $S_A=2\delta S^0_{RL}(x,-x,\omega)-\delta S^0(x,x,\omega)$ and
$S_C=\delta S^0(x,x,\omega)-\delta S^0_{RL}(x,-x,\omega)$. Similar combinations of the emission and the absorption noise could also be used to separate the two terms. The individual measurements of $S_A$ and $S_C$ may allow one to extract important information about the system; for example we note that for non-interacting systems $S_C$ is voltage-independent; the fact that
$S_C$ depends on voltage only in the presence of interactions could provide a direct way to reveal
them in a FQHE system.

\subsection{Generalized Kubo formula}

It has been shown that one can also write a generalization of the Kubo formula for non-linear systems in the linear regime corresponding to small values of the applied voltage. In this regime the antisymmetric part of the current noise can be related to the differential AC conductivity such that
\be
S(-\omega)-S(\omega)=2 \hbar \omega G_d(\omega)
\ee
where $G_d(\omega)$ is the differential conductivity of the system.
We can derive a more general relation for our system by analyzing the antisymmetric part of the noise.
We start from the general position-dependent non-perturbative results presented in Eqs.(\ref{seq},\ref{n1},\ref{n2},\ref{n3}), and we analyze the connection between the noise and the Green's functions of the system modified to take into account the effect of local backscattering. Thus, we can rewrite Eqs.(\ref{seq},\ref{n1},\ref{n2},\ref{n3}) such that
the total symmetrized current fluctuations are related to the
Keldysh Green's function  $\tilde{\bf C}^{\cal K}$ in the presence of backscattering:
\be
S^0(x,y,\omega)=\frac{e^2\omega^2}\pi \tilde{\bf C}^{\cal K}(x,y,\omega)
\label{se}
\ee
while the non-symmetrized noise is related to $\tilde{\bf C}^{+-}$:
\be
S^1(x,y,\omega)=\frac{e^2\omega^2}\pi \tilde{\bf C}^{+-}(x,y,\omega)
\label{sn}
\ee
where consistent with our previous notations, the indices $0$ and $1$ denote the symmetrized/non-symmetrized noises respectively.

Here we can see that Eqs.(\ref{seq},\ref{n1},\ref{n2},\ref{n3}) imply that the Green's functions $\tilde{\bf C}^{\cal K}$ and $\tilde{\bf C}^{{\cal R},{\cal A}}$ obey Dyson-type equations with the roles of self-energy being played by the
functions  $f_A(\omega)$ and $f_C(\omega)$.

\ba
e^2\tilde{\bf C}^{\cal R}(x,y,\omega)&=&e^2 \tilde{\cal C}^{\cal R}(x,y,\omega)-\tilde{\cal C}^{\cal R}(x,0,\omega) f_c(\omega)
\tilde{\cal C}^{\cal R}(0,y,\omega)\\
\tilde{\bf C}^{\cal A}(x,y,\omega)&=&\tilde{\bf C}^{\cal R}(y,x,-\omega)\\
e^2\tilde{\bf C}^{\cal K}(x,y,\omega)&=& e^2\tilde{\cal C}^{\cal K}(x,y,\omega)-\tilde{\cal C}^{\cal R}(x,0,\omega)f_A(\omega)
\tilde{\cal C}^{\cal A}(0,y,\omega)
\nonumber \\
&&-\tilde{\cal C}^{\cal R}(x,0,\omega) f_C(\omega) \tilde{\cal C}^{\cal K}(0,y,\omega)
-\tilde{\cal C}^{\cal K}(x,0,\omega) f_C(-\omega) \tilde{\cal C}^{\cal A}(0,y,\omega)
\ea

Noting that by the definition of the Keldysh transformation
\begin{equation}
2 \tilde{\bf C}^{+-}=\tilde{\bf C}^{\cal K}+\tilde{\bf C}^{\cal A}-
\tilde{\bf C}^{\cal R}
\end{equation}
we can see that the difference between the symmetrized and the non-symmetrized noises
comes from $\tilde{\bf C}^{\cal A}-
\tilde{\bf C}^{\cal R}$. Notice that a similar behavior will be observed if one works with
the chiral Green's function instead of the total Green's functions.
Thus for the fluctuations of the current evaluated at the same position in space we can write:
\ba
&&S^1(x,x,\omega)-S^1(x,x,-\omega)=\frac{e^2 \omega^2}{\pi}[
\tilde{\bf C}^{+-}(x,x,\omega)-\tilde{\bf C}^{+-}(x,x,-\omega)]
\nonumber \\&&
=\tilde{\bf C}^{\cal A}(x,x,\omega)-\tilde{\bf C}^{\cal A}(x,x,-\omega)-\tilde{\bf C}^{\cal R}(x,x,\omega)+\tilde{\bf C}^{\cal R}(x,x,-\omega)
\ea
We can also show that
$\tilde{\bf C}^A(x,x,-\omega)$ $=\tilde{\bf C}^R(x,x,\omega)$ $=-\tilde{\bf C}^R(x,x,-\omega)^*$, and we can rewrite the above
$exact$ expression as
\ba
S^1(x,x,-\omega)-S^1(x,x,\omega)=\frac{4e^2\omega^2}{\pi} {\rm{Re}}[
\tilde{\bf C}^{\cal R}(x,x,\omega)].
\ea
If we use the definition of the retarded Green's function presented in Eq.(\ref{cret}), and the definition of the current in Eq.(\ref{current}), we find $ \omega^2 \tilde{\bf C}^{\cal R}(x,x,\omega)$  to be proportional to the Fourier transform of the non-linear response function of the system $\theta(t-t')
\langle [j_a(\mathbf{x,t}), j_a(\mathbf{x,t'}) ] \rangle$.

In the limit $e V\ll \hbar \omega$ we can relate the non-local AC conductivity and the retarded Green's function by
\be
\sigma(x,y,\omega)=\frac{2e^2\omega}{h}\tilde{\bf C}^{\cal R}(x,y,\omega)
\label{sig}
\ee
and we find that
\be
S^1(x,x,-\omega)-S^1(x,x,\omega)=4 \hbar \omega {\rm{Re}}[
\sigma(x,x,\omega)].
\ee
This extends the generalized Kubo formula presented in \cite{kubo}. The extra factor of $2$ is due to our definition of noise.

We can now also verify explicitly how our perturbative
results satisfy this condition. For clarity we focus on the case $\omega x\ll 1$  and we drop all spatial indices.
We find:
\ba
S^1(-\omega)-S^1(\omega)&=&s^1(-\omega)-s^1(\omega)
+\delta S^1(-\omega)-\delta S^1(\omega)
\nonumber \\
&=&\frac{2 g e^2 \omega}{\pi}-2 g e \sum_{m=\pm1} I_B(\omega+m \omega_0)
\ea
We note that for small applied voltages this can be expanded as
\be
S^1(-\omega)-S^1(\omega)=4 \hbar \omega \sigma_0(\omega)-4 g e I_B(\omega)
\ee
where $\sigma_0(\omega)=g e^2/h$ is the unperturbed conductivity of the system.

Following Ref. \cite{dolcini} we define the conductivity of the system to be
\begin{equation}
\sigma(x,y,\omega)=\sigma_0(x,y,\omega)+ \sigma_{\rm
BS}(x,y,\omega) \label{sig12}
\end{equation}
where $\sigma_0(x,y,\omega)=g e^2/h$ is the conductivity of the
system in the absence of the backscattering, and
\begin{eqnarray}\nonumber
\sigma_{\rm BS}(x,y,\omega)= - \frac{2}{\hbar \omega} \left(
\frac{\pi \lambda}{e}\right)^2 \sigma_0(x,0,\omega)
\sigma_0(0,y,\omega) \\
\label{sigbs} \times \int_0^\infty dt \, (e^{i \omega t}-1) \left(
\sum_{s=\pm} s \, e^{4 \pi {\cal C}(0,s t;0,0)} \right)
\label{sg}
\end{eqnarray}
is the contribution to the conductivity due to backscattering. Here $\cal C$
is the generalized two-point function for an infinite Luttinger liquid in the absence of backscattering given in Eq.(\ref{Creg_def}) of Appendix A. The backscattered current can be expressed in terms of this conductivity as
\be
{\rm{Re}}[\sigma_{\rm BS}(\omega)]=-\frac{ge}{\hbar \omega} I_B(\omega)
\ee
Thus we can write
\be
S^1(-\omega)-S^1(\omega)=4 \hbar \omega \{\sigma_0+{\rm{Re}}[\sigma_{\rm BS}(\omega)]\}
\ee
which is consistent with the generalized Kubo formula presented above.

\subsection{Fluctuation dissipation theorem}
We check that the noise at zero voltage can indeed be related by the fluctuation-dissipation theorem to the real part of the conductivity.
The noise is described by Eqs.(\ref{se},\ref{sn}) presented in the previous subsection. At zero voltage we expect that
\be
\tilde{\bf C}^{\cal K}(x,y,\omega)=2  \coth\Big(\frac{\hbar \omega}{2 k_B T}\Big)
{\rm{Re}}[\tilde{\bf C}^{\cal R}(x,y,\omega)]
\ee
In the absence of backscattering, we see from Eq.(\ref{gf}) that this is indeed the case. Using the relation between the retarded Green's function and the conductivity in Eq.(\ref{sig}), we then expect
\be
S^{\alpha}(x,x,\omega)=2 \hbar \omega \Big[\coth\Big(\frac{\hbar \omega}{2 k_B T}\Big)-\alpha\Big] {\rm Re}[\sigma(x,x,\omega)].
\ee

We can check how this relation is satisfied by our perturbative results. We found that the noise in the absence of an applied voltage is:
\be
S^{\alpha}(x,x,\omega)=\Big[\coth\Big(\frac{\hbar \omega}{2 k_B T}\Big)-\alpha\Big]\Big[\frac{g \omega e^2}{\pi} + 2 g e  I_B(\omega)\Big]
\ee
which can be rewritten as
\be
S^{\alpha}(x,x,\omega)=2 \hbar \omega \Big[\coth\Big(\frac{\hbar \omega}{2 k_B T}\Big)-\alpha\Big] \{\sigma_0(x,x,\omega)+{\rm Re}[\sigma_{\rm BS}(x,x,\omega)]\}.
\ee
This is indeed the fluctuation-dissipation theorem for the symmetrized and for the non-symmetrized noise.

\subsection{Tunneling limit}
\begin{figure}[htbp]
\begin{center}
\includegraphics[width=3in]{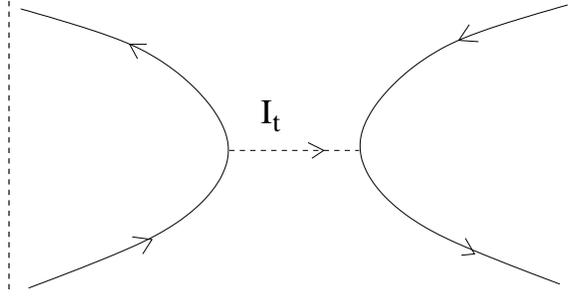}
\vspace{0.15in} \caption{\small Schematics of the sample for the tunneling regime}
\label{samplet}
\end{center}
\end{figure}

For completion we can also analyze the noise in the limit when the magnitude of the backscattering is very large. In this limit we can treat the system as being disconnected into two subsystems by the scattering site, with only a small amount of tunneling between the two subsystems. This limit is not so relevant for measuring the fractional charge of the quasiparticles, as in this situation the shot noise contains primarily information about the electrons tunneling between the two subsystems. We thus expect the ratio of the noise at zero frequency to the tunneling current to be proportional to $e$ instead of $g e$ \cite{kf}. Also, the Josephson frequency is shifted to $e V/\hbar$ instead of $g e V\hbar$. The total noise in this limit is given solely by a single term describing the fluctuations in the tunneling current between the two edges. For simplicity, we restrict ourselves to the limit $\omega x/v_F \ll1 1$ and we find the noise to be:

\ba
S_t^{\alpha}(\omega)&=&\frac{1}{4 \pi}\tilde{f}_A(\omega)-\alpha \frac{1}{4 \pi} [\tilde{f}_C(\omega)-\tilde{f}_C(-\omega)]
\nonumber \\
&=& e \sum_{m=\pm 1}\coth\Big[\frac{\hbar (\omega+m \tilde{\omega}_0)}{2 k_B T}-\alpha\Big]I_t(\omega+m  \tilde{\omega}_0)
\ea
with $\tilde{\omega}_0=e V/\hbar$, and
\begin{equation}
\tilde{f}_A(\omega) = \int_{-\infty}^\infty dt \, e^{i \omega t}
\left\langle \left\{ \hat{I}_t(t), \hat{I}_t(0) \right\}
\right\rangle
\end{equation}
where $\hat{I}_t$ is the tunneling current operator between the two edges.
Similarly
\begin{equation}
\tilde{f}_C(\omega) = \int_0^\infty dt  e^{i \omega t}
\left\langle \left[ \hat{I}_t(t),\hat{I}_t(0) \right]
\right\rangle
\end{equation}
Also
\be
I_t(\omega)= - \frac{e}{\hbar^2} |\Gamma|^2  {\cal{F}}_{1/g}(\omega)
\ee
is the value of the tunneling current for an applied voltage equal to $\omega$, with
\ba
{\cal F}_{1/g}(\omega)=i \sin\Big(\frac{\pi}{g}\Big) \Big(\frac{\pi k_B T}{\epsilon_h}\Big)^{2/g} 2^{2/g-2} \Gamma(1-2/g)\Big[\frac{\Gamma(1/g-i \tilde{\omega})}{\Gamma(1-1/g-i \tilde{\omega})}-\frac{\Gamma(1/g+i \tilde{\omega})}{\Gamma(1-1/g+i \tilde{\omega})}\Big]
\ea
where $\tilde{\omega}=\hbar \omega/2 \pi k_B T$.

\begin{figure}[htbp]
\begin{center}
\includegraphics[width=3in]{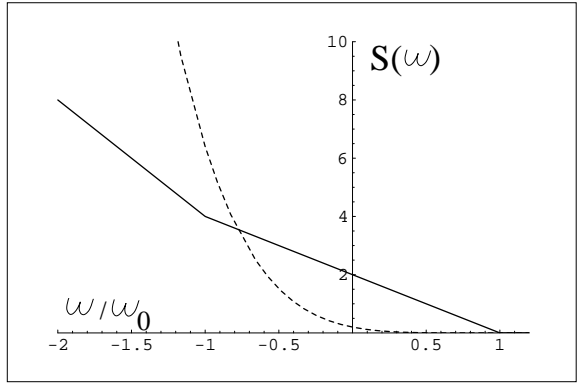}
\vspace{0.15in} \caption{\small The non-symmetrized noise in the tunneling current $S_t(\omega)$ as a function of frequency (in units of $\omega_0=e V/\hbar$) for $\nu=1/3$ (full line), and $\nu=1$ (dashed line). The noise $S_t(\omega)$ is given in units of $e I_t$ for the case of $g=1$, while for $g=1/3$ the units are arbitrary to facilitate the plot of the two curves on the same figure without a loss of information at large negative frequencies.}
\label{tunn}
\end{center}
\end{figure}

We see that indeed at zero frequency, the symmetrized noise $S^0(\omega \rightarrow 0)=2 e I_t(\tilde{\omega}_0)$, while at high frequency the noise has a singularity at a Josephson frequency $\tilde{\omega}_0$ given by $e V/\hbar$. Also, information about the Luttinger liquid characteristics of the FQHE edges is contained in the $1/g$ exponent, which is the dual of the $g$ exponent in the case of small backscattering between the two edges.
When  we plot the non-symmetrized noise as a function of frequency for a few values of $g$ (see Fig.~\ref{tunn}), we find that indeed the case of $g=1$ is consistent with the known results for non-interacting fermions. The case of small $g$ shows that indeed the noise has a power-law dependence $\omega^{2/g-1}$ on frequency; however the singularities at the Josephson frequency $\tilde{\omega}_0$ are very soft in this situation, for example the singularity in the emission noise is of the type  $\theta(\tilde{\omega}_0-\omega)|\omega-\tilde{\omega}_0|^5$ for $g=1/3$ which is very weak and becomes basically impossible to observe. A similar situation happens for the absorption noise, as can be seen from Fig.~\ref{tunn}.

\section{Conclusion}
We computed the high frequency non-symmetrized noise for a FQHE sample with small backscattering. At zero frequency we retrieved the classical result, $S=2 g e I_B$. At finite frequency and zero temperature the most striking feature we observe is a singularity at the Josephson frequency $\omega_0=g e V/\hbar$. This singularity is rounded off by temperature.  For a non-interacting system ($\nu=1$) this feature is cusp-like, while it has power-law resonant characteristics (a peak, or a ``positive peak - negative dip'' structure depending on the quantity measured) in a strongly interacting system, e.g. ($\nu=1/3$).

Other important aspects that we observed were that the
non-symmetrized correlations of chiral outgoing branches are
entirely even in frequency. However, the backscattering-induced
noise in the total current is not even. If one looks instead
at the excess noise we see that this is even in the absence
of interactions, but it becomes asymmetric if the
electron-electron interactions are taken into account. Thus, the
asymmetry with respect to positive and negative frequencies
in the excess noise can be used to assess the Luttinger
liquid character of a system. Also we noted that the
excess noise (both symmetrized and non-symmetrized) can become
negative in some regions of frequencies, and for
very strong electron-electron interactions; this may provide a
signature of Luttinger liquid physics.

If the distance between the scattering site and the measuring point is significant, oscillations of the noise with respect to frequency will also appear. The period of these oscillations is proportional to the fractional charge $g$. Such oscillations do not occur in correlations of chiral outgoing branches, but are manifest in the absorption part of the non-symmetrized noise of the total current, as well as in the symmetrized noise of the total current.

We have also analyzed also the limit when the strength of backscattering is very large and the system
is disconnected in two subsystems, in this situation the noise
is dominated by electrons tunneling between the two subsystems.

We should note that, while not done here, for the case of weak backscattering it would also be interesting to compute the higher order corrections in the backscattering amplitude $\Gamma$ for the emission and absorption noises at high frequencies.
As shown in \cite{chetan}, the fourth order contributions  to the high frequency symmetrized noise in a Laughlin state as well as in a non-Abelian Pfaffian state contain information about fractional and non-Abelian statistics; it would also be interesting to see what are the effects of the higher order corrections for the non-symmetrized noise.

\vspace{.2in}

\noindent{\bf\large Acknowledgements:}~
We would like to thank Richard Deblock, Christian Glattli, Frank Hekking, Fabien Portier, Patrice Roche, Bertrand Reulet and Hubert Saleur for interesting discussions. CB acknowledges the support of a Marie Curie Intra-European Fellowship.

\vspace{.4in}

{\noindent{\Large{\bf Appendix A: The Keldysh formalism used to calculate the non-symmetrized noise}}}
\vspace{.1in}

\begin{figure}[htbp]
\begin{center}
\includegraphics[width=2.5in]{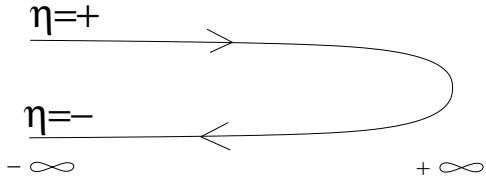}
\vspace{0.15in} \caption{\small The two branches of the Keldysh contour}
\label{pfaf}
\end{center}
\end{figure}

As depicted in Fig.\ref{pfaf}, we introduce the standard Keldysh time contour \cite{Keldysh} and we denote by $\Phi^+_a$ and $\Phi^-_a$ ($a=R/L$ denotes the chirality) the complex fields on
the upper and lower time branch of the contour. We
introduce the generating functional
\begin{eqnarray}
&&Z[J]= \frac{1}{\cal{N}_Z} \int {\mathcal D}
\Phi^{\pm}_{R/L}  \exp \Big\{-\frac{1}{2} \int d\mathbf{r'}
d\mathbf{r''} \sum_{\eta,\eta'=\pm} \sum_{a=R/L}\Phi^{\eta}_a(\mathbf{r'})
({\mathcal C}^{-1})^{\eta,\eta'}_{a}
(\mathbf{r'},\mathbf{r''})
\Phi^{\eta'}_a(\mathbf{r''}) \Big\} \nonumber
\\ && \times   \exp\Big\{\sum_{\eta=\pm} \Big( -\frac{i}{\hbar} \eta
\int_{-\infty}^{+\infty} d t' \mathcal{H}_B[\Phi^{\eta}]\Big) +\sum_{\eta=\pm,a=R/L}i \int
d\mathbf{x} J_a^{\eta}(\mathbf{x})\Phi_a^{\eta}(\mathbf{x})\Big)\Big\}
\label{gen-fun-1}
\end{eqnarray}
where  the vector label $\mathbf{r'}$ stands for
$\mathbf{r'}=(x',t')$, $\int d\mathbf{r'}=\int_{-\infty}^{+\infty}
dt' \int_{-\infty}^{+\infty} dx'$, and $\cal{N}_Z$ is a
normalization factor, which assures that $Z[0]=1$. In
Eq.~(\ref{gen-fun-1}), ${\mathcal
C}^{-1}(\mathbf{r'},\mathbf{r''})$ is the inverse of a $4 \times
4$ matrix defined by the four free correlators
\begin{equation}
{\mathcal C}^{\eta,\eta'}_{a a}(\mathbf{r'};\mathbf{r''})={\mathcal C}^{\eta,\eta'}_{a}(\mathbf{r'};\mathbf{r''}) = \langle
\Phi^\eta_a(\mathbf{r'}) \Phi^{\eta'}_a(\mathbf{r''}) \rangle_0
\end{equation}
and
\begin{equation}
{\mathcal C}^{\eta,\eta'}_{a b}(\mathbf{r'};\mathbf{r''})=0
\end{equation}
if $a \ne b$.
Also, $\langle \ldots \rangle_0 $ indicates the average performed
with respect to the free Hamiltonian (\ref{l0}) along the Keldysh
contour.
For a Luttinger liquid with a backscattering site and in presence of an applied voltage,
we have
\begin{equation}
\langle \Phi_a(\mathbf{x}) \rangle = \frac{1}{2} \sum_{\eta=\pm}
\langle \Phi^\eta_a(\mathbf{x}) \rangle =-\frac{i}{2} \sum_{\eta=\pm}\left.
\frac{\delta Z[J]}{\delta J_a^{\eta}(\mathbf{x})} \right|_{J=0}\;.
\label{Phi-VM1}
\end{equation}
One can simplify the notation by introducing infinite-dimensional
vectors and matrices where {\it both} $\mathbf{r}$, $\eta$ and $a$ are
component labels. Defining
\begin{equation}
\mathbf{\Phi}= \left(
\begin{array}{c} \Phi^{+}_R(\mathbf{r}) \\ \Phi^{-}_R(\mathbf{r})\\ \Phi^{+}_L(\mathbf{r}) \\\Phi^{-}_L(\mathbf{r}) \end{array}
\right),
\end{equation}
\begin{equation}
\mathbf{J}=   \left(
\begin{array}{r}
   J_R^+(\mathbf{r})\\
 J_R^-(\mathbf{r})\\J_L^+(\mathbf{r})\\
 J_L^-(\mathbf{r})
\end{array}
\right) ,
\end{equation}
and
\begin{equation}
\mathsf{C}= \left(
\begin{array}{cccr}
\mathcal{C}^{++}_{R}(\mathbf{r},\mathbf{r'}) &
\mathcal{C}^{+-}_{R}(\mathbf{r},
\mathbf{r'})&0&0 \\
\mathcal{C}^{-+}_{R}(\mathbf{r},\mathbf{r'}) &
\mathcal{C}^{--}_{R}(\mathbf{r},
\mathbf{r'}) &0&0\\
0&0&\mathcal{C}^{++}_{L}(\mathbf{r},\mathbf{r'}) &
\mathcal{C}^{+-}_{L}(\mathbf{r},
\mathbf{r'})\\
0&0&\mathcal{C}^{-+}_{L}(\mathbf{r},\mathbf{r'}) &
\mathcal{C}^{--}_{L}(\mathbf{r},
\mathbf{r'})\\
\end{array}
\right) ,
\end{equation}
one can rewrite the generating functional (\ref{gen-fun-1}) as
\begin{eqnarray}
\displaystyle Z[J]= \frac{1}{\cal{N}_Z} \int {\mathcal D}
\mathbf{\Phi} \, e^{-\frac{1}{2} \Big(\mathbf{\Phi}^T  {\mathsf
C}^{-1} \mathbf{\Phi} -2 i \mathbf{J}^T
\mathbf{\Phi} \Big)}
 \exp{\left\{ - \frac{i}{\hbar} \sum_{\eta=\pm} \eta
\int_{-\infty}^{+\infty} d t' \mathcal{H}_B[\Phi^{\eta}]
\right\}}\, ,
\end{eqnarray}
where the superscript ${}^T$ indicates the transpose. Shifting the
fields
\begin{equation}
\mathbf{\Phi} \rightarrow \mathbf{\Phi}+{\mathbf{A}}\, ,
\hspace{1cm} {\mathbf{A}}=i{\mathbf{\mathsf{C}}}
 \mathbf{J} \, , \label{shift-in-field}
\end{equation}
the generating functional can be factorized into
\begin{equation}
Z[J]=Z_0[J]  Z_B[J] \; , \label{fun-gen-fact}
\end{equation}
where $Z_0$ and $Z_B$ are given below. In particular, $Z_0$ is the
generating functional in the absence of a backscatterer and reads
\begin{equation}
Z_0[J]= e^{-\frac{1}{2} \mathbf{J}^T {\mathsf C}
\mathbf{J}} \label{Z0}
\end{equation}
The second factor $Z_B$ in (\ref{fun-gen-fact}) is the generating
functional
\begin{eqnarray}\label{ZB}
 Z_B[J(\mathbf{r})]
 = \left \langle    \exp{\left(  -
\frac{i}{\hbar}  \sum_{\eta=\pm} \eta \int_{-\infty}^{+\infty}
\mathcal{H}_B[\Phi^{\eta}_R+\Phi^{\eta}_L+A^{\eta}_R+A^{\eta}_L]  \, d t' \right)}
\right\rangle_0,
\end{eqnarray}
which  weighs the backscattering term, and where the dependence
on the source field $J(\mathbf{x})$ is contained in the shift
${\mathbf{A}}$ defined in Eq.~(\ref{shift-in-field}). In
components, the latter reads explicitly
\ba
&&{\mathbf{A}}_{a}^{\eta}=\int d \mathbf{x}
 i \sum_{\eta'=\pm} C^{\eta \eta'}_{a}(\mathbf{r};\mathbf{x}) J_a^{\eta'}(\mathbf{x})
\ea
The Keldysh Green's functions can be related to the regular Green's function using the
relations:
\be
C_a^{\eta_1 \eta_2}=\frac{\eta_1 C_a^{\cal A}+\eta_2 C_a^{\cal R}+C_a^{\cal K}}{2}
\ee
Here $C_a^{{\cal R}/{\cal A}}$ are the retarded and advanced Green's function respectively,  $\eta_{1,2}=\pm$, and $C_a^{{\cal K}}$  is defined in Appendix B; for an infinite Luttinger liquid we evaluate it explicitly in Appendix B.

The finite frequency non-symmetrized noise $S_{a b}(x,y,\omega)$ is defined as
\begin{eqnarray}
S_{a b}(x,y,\omega) = \int_{-\infty}^{\infty} dt e^{i\omega t}
2 \left\langle \Delta j_a (y,0)  \Delta j_b(x,t)
\right\rangle \; , \label{noiseeq}
\end{eqnarray}
where $a,b$ stand for the right/left moving indices, and $\Delta
j_a(x,t) = j_a(x,t) - \langle j_a(x,t) \rangle$ is the current
fluctuation operator, with $j_a(x,t)=e \partial_t \Phi_a(x,t)/\sqrt{\pi}$.
Thus, using the Keldysh formalism we can express the non-symmetrized noise as
\ba
S_{a b}(x,y,\omega)&=&\int_{-\infty}^{\infty} dt_x e^{i \omega t_x}\frac{2 e^2}{\pi} \partial_{t_x} \partial_{t_y} \langle \Phi_a(\mathbf{y})\Phi_b(\mathbf{x})\rangle |_{t_y=0}\\ &=&
-\frac{2 e^2}{\pi} \int_{-\infty}^{\infty} dt_x e^{i \omega t_x}\Big\{\partial_{t_x} \partial_{t_y}
 \frac{\partial^2 Z}{\partial J_a^+(\mathbf{x})\partial J_a^-(\mathbf{y})}\Big\}\Big|_{t_y=0}.
\ea
The noise in the total current is obtained by summing up $S_{RR}+S_{LL}+S_{RL}+S_{LR}$.
The procedure of taking the functional derivatives is lengthy but straightforward and follows closely the procedure described in Ref. \cite{dolcini}. We obtain
\be
S^{\alpha}_{a b}
(x,y,\omega)=s^{\alpha}_{a b}(x,y,\omega)+
S^A_{ a b}(x,y,\omega)+S^C_{a b}(x,y,\omega)+\alpha S^N_{a b}(x,y,\omega),
\ee
where $\alpha=0$ for the symmetrized noise and $\alpha=1$ for the non-symmetrized noise.
Here we have

\ba
s^{\alpha}_{a b}(x,y,\omega)&=&\delta_{a b}\frac{e^2 \omega^2}{\pi} \{\tilde{\cal C}^{\cal K}_{a}(x,y,\omega)+\alpha [\tilde{\cal C}^{\cal A}_a(x,y,\omega)-
\tilde{\cal C}_a^{\cal R}(x,y,\omega)]\},\\
S^A_{a b}(x,y,\omega)&=&-\frac{\omega^2}{\pi} \tilde{\cal C}^{\cal R}_{a}(x,0,\omega)
\tilde{\cal C}^{\cal R}_{b}(y,0,-\omega)f_A(\omega),\nonumber \\
S^C_{a b}(x,y,\omega)&=&-\frac{\omega^2}{\pi}[ \tilde{\cal C}^{\cal K}_{a}
(x,0,\omega)\tilde{\cal C}^{\cal R}_{b}(y,0,-\omega)f_C(-\omega)
\nonumber \\&&
+\tilde{\cal C}^{\cal R}_{a}(x,0,\omega)\tilde{\cal C}^{\cal K}_{b}
(y,0,-\omega)f_C(\omega)],\nonumber \\
S^N_{a b}(x,y,\omega)&=&-\frac{\omega^2}{\pi}[\tilde{\cal C}^{\cal A}_{a}(x,0,\omega)
\tilde{\cal C}^{\cal R}_b(y,0,-\omega)f_C(-\omega)\nonumber \\&&-
\tilde{\cal C}^{\cal R}_a(x,0,\omega)\tilde{\cal C}^{\cal A}_b(y,0,-\omega)f_C(\omega)]
\nonumber
\ea
with
\begin{equation}
\tilde{\mathcal{C}}^m(x,y,\omega)=\int_{-\infty}^{\infty} \,
e^{i \omega t} \, {\mathcal{C}}^m(x,t;y,0) \, dt \; ,
\end{equation}
for $m={\cal A,R,K}$, and $x_0$ is the backscattering position
which we will set to zero.

The functions $f_A(\omega)$ and $f_C(\omega)$ are given by
\begin{equation}
f_A(\omega) = \int_{-\infty}^\infty dt \, e^{i \omega t}
\left\langle \left\{ \Delta j_B(t), \Delta j_B(0) \right\}
\right\rangle , \label{s1_A}
\end{equation}
where $\Delta j_B(t) = j_B(t) - \langle j_B(t)
\rangle$. The expectation values are performed with respect to the full action.
Also, $j_B(t)$ is the backscattering
current operator at the backscattering site
\begin{equation}
j_{B}(t) = - \frac{e}{\hbar} \frac{\delta
\mathcal{H}_B(\Phi)}{\delta \Phi(0,t)} \; \label{ib1_def} .
\end{equation}
Similarly
\begin{equation}
f_C(\omega) = \int_0^\infty dt \left( e^{i \omega t}-1 \right)
\left\langle \left[ j_B(t),j_B(0) \right]
\right\rangle \; . \label{s1_C}
\end{equation}
We can evaluate $f_A$ and $f_C$ perturbatively up to second order in $\Gamma$ to find:
\ba
f_A(\omega)&=&
4 \pi \Big(\frac{e}{\hbar}\Big)^2 |\Gamma|^2 \int_0^\infty dt \cos(\omega t)
\cos(\omega_0 t) \sum_{s=\pm} e^{4 \pi {\cal C}(0,s t;0,0)} \\
&=& 2 \pi i \Big(\frac{e}{\hbar}\Big)^2 |\Gamma|^2 \sum_{m=\pm 1} \coth \Big[ \frac{\hbar(\omega+m \omega_0)}{2 k_B T}\Big]\int_0^\infty dt \sin[(\omega+m \omega_0) t] \sum_{s=\pm}s e^{4 \pi {\cal C}(0,s t;0,0)} \nonumber
\ea
and
\ba
f_C(\omega)=2 \pi \Big(\frac{e}{\hbar}\Big)^2 |\Gamma|^2 \int_0^\infty dt (e^{i \omega t}-1)
\cos(\omega_0 t) \sum_{s=\pm}s e^{4 \pi {\cal C}(0,s t;0,0)}
\ea
where $\omega_0=g e V/\hbar$, and ${\cal C}$ is the generalized
correlation function
\begin{eqnarray}
{\cal C}(x,t;y,0)= \left\langle \, \Phi(x,t) \Phi(y,0) -  \frac{\Phi^2(x,t) +
\Phi^2(y,0)}{2} \, \right\rangle_0 \label{Creg_def}.
\end{eqnarray}
with $\Phi=\Phi_R+\Phi_L$.

The Green's functions for an infinite Luttinger liquid system are presented in Appendix B. Using their specific forms we find that for  $x=y>0$, the noise in the absence of backscattering is
\ba
&&s^{\alpha}_{RR}(x,x,\omega)=s^{\alpha}_{L L}(x,x,\omega)=\frac{g \omega e^2}{2 \pi} \Big[\coth\Big(\frac{\hbar \omega}{2 k_B T}\Big)-\alpha\Big] \nonumber \\
&&s^{\alpha}(x,x,\omega)=\frac{g \omega e^2}{\pi}
\Big[\coth\Big(\frac{\hbar \omega}{2 k_B T}\Big)-\alpha\Big] \ea
with the cross correlations between the right and the left-movers
vanishing. The backscattering induced noise is \ba
&&\delta S^{\alpha}_{R R}(x,x,\omega)=\frac{g^2}{4\pi} f_A(\omega)- \frac{g^2}{4 \pi} \coth\Big(\frac{\hbar \omega}{2 k_B T}\Big)[f_C(\omega)-f_C(-\omega)]\nonumber \\
&&\delta S^{\alpha}_{LL}(x,x,\omega)=0\nonumber \\
&&\delta S^{\alpha}_{RL}(x,x,\omega)+\delta S^{\alpha}_{LR}(x,x,\omega)=
- \frac{g^2}{4 \pi} \Big[\coth\Big(\frac{\hbar \omega}{2 k_B T}\Big)-\alpha\Big]\times
\nonumber \\&&
\times \Big\{\exp\Big(\frac{2 i g  \omega |x|}{v_F}\Big) f_C(\omega)
- \exp\Big(-\frac{2 i g  \omega |x|}{v_F}\Big) f_C(-\omega)\Big\}
\label{qet}
\ea
Here, as before, we denote the symmetrized noise by $\alpha=0$ and the non-symmetrized noise by $\alpha=1$.
The noise in the total current is
\ba
&&\delta S^{\alpha}(x,x,\omega)=
\frac{g^2}{4\pi} f_A(\omega)- \frac{g^2}{4 \pi} \coth\Big(\frac{\hbar \omega}{2 k_B T}\Big)[f_C(\omega)-f_C(-\omega)]
- \frac{g^2}{4 \pi} \Big[\coth\Big(\frac{\hbar \omega}{2 k_B T}\Big)-\alpha\Big]
\times
\nonumber \\&& \times \Big\{\exp\Big( \frac{2 i g  \omega |x|}{v_F}\Big) f_C(\omega)
- \exp\Big(-\frac{2 i g  \omega |x|}{v_F}\Big) f_C(-\omega)\Big\}
\ea
Similarly, we find:
\ba
&&\delta S^{\alpha}_{RL}(x,-x,\omega)=\frac{g^2}{4\pi} f_A(\omega)- \frac{g^2}{4 \pi} \coth\Big(\frac{\hbar \omega}{2 k_B T}\Big)[f_C(\omega)-f_C(-\omega)]
\nonumber \\&&
\delta S^{\alpha}_{R R/L L}(x,-x,\omega)= \mp \frac{g^2}{4 \pi}\Big[\coth\Big(\frac{\hbar \omega}{2 k_B T}\Big)-\alpha\Big]f_C(\pm \omega) e^{\pm 2 i g  \omega x/v_F}
\ea

Noting that
\be
f_A(\omega)=- 4 \pi \frac{e^2}{\hbar^2} |\Gamma|^2 \sum_{m=\pm 1}\coth\Big[\frac{\hbar (\omega+m \omega_0)}{2 k_B T}\Big]{\cal F}_g(\omega+m  \omega_0)
\ee
and
\be
f_C(\omega)-f_C(-\omega)=- 4 \pi \frac{e^2}{\hbar^2} |\Gamma|^2 \sum_{m=\pm 1}{\cal F}_g(\omega+m  \omega_0)
\ee
where ${\cal F}_g(\omega)$ was defined in Eq. (\ref{calf}), we retrieve the equations presented in section 3.
\vspace{.2in}

{\noindent{\Large{\bf Appendix B  - Chiral Green's functions for an infinite Luttinger liquid}}}

\vspace{.1in}
The correlations for two chiral operators in a Luttinger liquid can be computed from the fundamental correlations:
\begin{eqnarray}
{\cal C}_{a}(x,t;y,0)= \left\langle \, \Phi_a(x,t) \Phi_a(y,0) -  \frac{\Phi^2_a(x,t) +
\Phi^2_a(y,0)}{2} \, \right\rangle_0 \label{Creg_def2}
\end{eqnarray}
We find that
\be
{\cal C}_{R/L}(x,t;y,0)=-\frac{g}{4 \pi} \log \Big\{\frac{1+i[t\mp g(x-y)/v_F]\omega_h}{1\mp i g(x-y)/v_F \omega_h}\Big\}+\frac{g}{4 \pi} \log
\Big\{\frac{\pi T[t \mp g(x-y)/v_F]}{\sinh \pi T [t \mp g(x-y)/v_F]}\Big\}
\ee
with the two-point function between a right mover and a left mover being zero. The total two-point function ${\cal C}$ is obtained by summing the two components.
By definition:
\begin{eqnarray}
{\mathcal C}_{a}^{\cal A}(\mathbf{r};\mathbf{r'}) &=& - \theta(t'-t)
\langle [ \Phi_a(\mathbf{r}), \Phi_a(\mathbf{r'}) ] \rangle_0 \; , \label{cadv} \\
{\mathcal C}_{a}^{\cal R}(\mathbf{r};\mathbf{r'}) &=& \theta(t-t')
\langle [ \Phi_a(\mathbf{r}), \Phi_a(\mathbf{r'}) ] \rangle_0 \; ,
\label{cret}
\\ {\mathcal C}_{a}^{\cal K}(\mathbf{r};\mathbf{r'}) &=& \langle \{
\Phi_a(\mathbf{r}),\Phi_a(\mathbf{r'})\} \rangle_0 \; \label{ckel}
\end{eqnarray}
with the corresponding Fourier transforms:
\begin{equation}
\tilde{\mathcal{C}}^m(x,y,\omega)=\int_{-\infty}^{\infty} \,
e^{i \omega t} \, {\mathcal{C}}^m(x,t;y,0) \, dt \; ,
\end{equation}
for $m={\cal A,R,K}$.

We obtain
\ba
&&\tilde{\mathcal{C}}^{\cal A}_{L/R}(x,y,\omega)=-\frac{g}{2 \omega}e^{\mp i g \omega (x-y)/v_F} \theta[\pm (x-y)] \nonumber \\&&
\tilde{\mathcal{C}}^{\cal A}(x,y,\omega)=-\frac{g}{2 \omega}e^{-i g \omega |x-y|/v_F}
\ea
and
\ba
&&\tilde{\mathcal{C}}^{\cal R}_{L/R}(x,y,\omega)=\frac{g}{2 \omega}e^{\mp i g \omega (x-y)/v_F} \theta[\mp (x-y)] \nonumber \\&&
\tilde{\mathcal{C}}^{\cal R}(x,y,\omega)=\frac{g}{2 \omega}e^{i g \omega |x-y|/v_F}
\ea
while, up to some constants
\ba
&&\tilde{\mathcal{C}}^{\cal K}_{L/R}(x,y,\omega)=\frac{g}{2 \omega}e^{\mp i g \omega (x-y)/v_F} \coth\Big(\frac{\hbar \omega}{2 k_B T}\Big) \nonumber \\&&
\tilde{\mathcal{C}}^{\cal K}(x,y,\omega)=\frac{g}{\omega} \cos[\omega(x-y)/v_F] \coth\Big(\frac{\hbar \omega}{2 k_B T}\Big).
\ea
We observe that the total $\cal C$ satisfies the fluctuation-dissipation theorem, while this is not the case for the individual chiral Green's functions.

\end{document}